\documentclass[11pt]{article}
\usepackage[title]{appendix}
\usepackage{amssymb}
\usepackage{mathtools}
\usepackage{epsf}
\usepackage{caption}
\usepackage{amsmath}
\usepackage{amsthm}
\usepackage{wrapfig}
\usepackage{natbib}
\usepackage{appendix}
\usepackage{enumitem}
\usepackage{graphicx}
\usepackage{hyperref}
\hypersetup{colorlinks, allcolors=blue}

\newcommand{\oP}{{\mathbb P}}
\newcommand{\oE}{{\mathbb E}}

\usepackage{inputenc}
\usepackage{tikz}
\usetikzlibrary{shapes.geometric, arrows}
\tikzstyle{startstop} = [rectangle, rounded corners, minimum width=3cm, minimum height=1cm,text centered, draw=black, fill=white]
\tikzstyle{process} = [rectangle, minimum width=2cm, minimum height=1cm, text centered, draw=black, fill=white]
\tikzstyle{decision} = [ellipse, minimum width=2cm, minimum height=1cm, text centered, draw=black, fill=white]
\tikzstyle{arrow} = [thick,->,>=stealth]

\begin{document}
\thispagestyle{empty}

\begin{center}
{\bf \large Incorporation of NAM's dynamic reservoir model into Cox rate-and-state model for monitoring of earthquakes in the Groningen gas field}\\[0.2in]
Z. Baki$\mbox{}^{a}$\\[0.2in]

$\mbox{}^a$
Faculty of Electrical Engineering, Mathematics and Computer Science, 
University of Twente, P.O.~Box 217, NL-7500 AE, Enschede, The Netherlands.
\end{center}
\bigskip
\noindent
\textbf{Abstract:} Induced seismicity due to fluid extraction or injection has become a critical issue in regions with extensive hydrocarbon production, such as the Groningen gas field in the Netherlands. This study examines the relationship between pore pressure changes from natural gas extraction and the resulting induced earthquakes in Groningen. We employ a Cox process-based rate-state model, integrating pore pressure data from NAM's dynamic reservoir model. By combining geomechanical and statistical approaches, we aim to predict future seismic events and assess the accuracy of our model. Our methodology uses Markov Chain Monte Carlo (MCMC) algorithms to estimate model parameters and forecast earthquake occurrences. The results highlight the significant impact of pressure changes on seismic activity, providing valuable insights for mitigating seismic risks in gas-producing regions.


\noindent \textbf{Mathematics Subject Classification (MSC 2020):} 60G55, 62F15, 62M30.

\section{Introduction}

The topic of induced seismicity resulting from fluid extraction and/or injection has become crucial over the years as the number of such instances increase. The Netherlands is no exception in this regard, as it has experienced a myriad of induced earthquakes from the Groningen gas field since the late 1980s. It is believed that the earthquakes in the northern Netherlands were caused by a drop in pore pressure due to natural gas extraction. The decrease in pore pressure results in the compaction of the reservoir, which in turn increases stress in the faults and results in induced seismicity. Thus, careful study of the geological properties and seismicity in the gas-producing regions is needed to improve the production process and minimize the seismic hazard.       

The extensive gas production in the Netherlands started in the early 1960s with the discovery of the Groningen gas field in the late 1950s. Initially, it boosted the Dutch economy greatly, with annual extraction volumes reaching up to 87 billion normal cubic meters (bcm). However, the increased number of induced earthquakes, the magnitude of which reached the $3.6$ Richter scale by 2012, forced the Dutch Ministry of Economic Affairs to reevaluate their plans and gradually decrease the extraction volumes. Finally, in October of 2023, the gas field was closed. According to data from the Dutch Oil and Gas Company (NAM), a total of 2249 bcm out of available 2900 bcm was extracted throughout the field's lifetime.     

Numerous studies have been conducted on induced seismicity due to fluid extraction and/or injection. Most notably, \citet{Broc17} proposed a hierarchical Bayesian model for fluid-induced seismicity based on the Basel 2006 study. \citet{Haja15} characterised injection-induced seismicity in terms of temporal Poisson processes. His assumptions were further expanded by \citet{Sija17}, who introduced a Bayesian method with change point analysis for the seismicity in Groningen. More recently, \citet{Post21} and \citet{Tram22} modelled interevent-time distributions of induced earthquakes. Post's model addressed spatio-temporal variations, while Trampert showed that interevent-time distributions remain scale-invariant when increasing the magnitude threshold for earthquakes. Several other researchers have focused their studies on the Groningen gas field, including but not limited to an entire special issue in the Netherlands Journal of Geosciences.       

Among one of the most notable state-of-the-art earthquake modelling methods is the rate-state model \citep{Cand19, DempSuck17, Rich20}. It assumes the earthquakes to be Poisson distributed with intensity $\lambda$ that is inversely proportional to a state variable $\Gamma$:
\begin{align*}
    d\Gamma(s,t) = \alpha( dt + \Gamma(s,t) dP(s,t))
\end{align*}
for pore pressure $P()$ and model parameter $\alpha$. In our previous work \citep{BakiLies24}, we have shown that discretizing time in steps of $\Delta$ the state variable can be rewritten in Euler equation form:
\begin{align*}
    \Gamma(s,t_0+k\Delta) & = \gamma_0\exp\{ \alpha [P(s,t_0+k\Delta) - P(s,t_0)]\} \\ & + \alpha\Delta \sum_{i=0}^{k-1} \exp\{ \alpha [ P(s, t_0 +k\Delta) - P(s, t_0 +i\Delta)]\}.
\end{align*}
for the starting time $t_0$, and initial state $\Gamma(s,t_0) = \gamma_0$. Then, we proposed to consider the earthquake occurrences $\Phi$ as a point process with intensity measure 
\begin{align*}
    \Lambda(s, t_0 + k\Delta) = \frac{\gamma_0\exp\{ \theta_1 + \theta_2 V(s,t_0+k\Delta)\}}{\Gamma(s,t_0+k\Delta)}
\end{align*}
for a volume of gas produced within a year leading to time $t_0+k\Delta$, $V(s, t_0+k\Delta)$. However, pore pressure observations are not universally available across the gas field, and some form of estimation is needed. We assumed the pore pressure values to be $P(s,t) = m(s,t) + E(s,t)$ for some deterministic function $m(s,t)$ such that $m(s,t) = \oE\{P(s,t)\}$ and zero-mean Gaussian distributed random field $E(s,t)$ with variance $\sigma^2$. Thus making the process $\Phi$ a spatio-temporal Cox process \citep{Cox55}. We then fitted a fourth-order polynomial to the observed pore pressure data and used it as a function $m(s,t)$. Additionally, we have developed a methodology for fitting the parameters and predicting future intervals using Markov Chain Monte Carlo (MCMC) algorithms. In this paper, we propose to use the developed methodology with the output pressure estimates from NAM's reservoir model \citep{NAMJan12}.   

A reservoir model is the computer representation of the subsurface containing water or hydrocarbon reservoir used to study the reservoir properties. The aim of such models is to predict the volume of fluid available and its distribution. NAM's reservoir model has two parts: static and dynamic \citep{NJG17_2}. A static model is based on geological concepts and describes fluid distribution in the reservoir and existing barriers \citep{NJG17_2}. Some examples include the presence of dissecting faults, the fluid composition in the pores, and physical characteristics like porosity, permeability, etc. Although these characteristics result from thousands of years of sedimentary formations in the context of fluid production, they are considered constant over time. On the other hand, the dynamic model uses the static model as a framework to describe the transport process of pore fluid throughout the reservoir. In hydrocarbon production, the dynamic models assess the depletion of pore pressure in the reservoir as a result of fluid extraction \citep{NJG17_2}. Common examples of dynamic models are pressure in the reservoir and subsurface subsidence. Since the production volumes and depletion amounts vary throughout the reservoir's lifetime, so do the parameters of the dynamic model. Thus, regular measurements throughout the field are performed, and the model is constructed to match the observed data closely. In the reservoir modelling context, this process is called history matching.

In this study, we propose integrating the pore pressure estimates generated by thus using the reservoir model's outputted pore pressure estimates and fitting them into the previously developed Cox rate-state model framework. By doing so, we combine a purely statistical model with the geomechanical one. We estimate future intervals and assess their accuracy. Additionally, we compare the performance of the two approaches and draw conclusions about their effectiveness.

\section{Data}\label{S:data}
In the Groningen gas field, there are three primary sources of data: the Dutch Oil and Gas Company (NAM), the Royal Dutch Meteorological Office (KNMI) and the Geological Survey of the Netherlands. Each collects and stores data on their perspective domain. 

TNO has data on the geological coordinates of the gas field \citep{TNO}. KNMI records and stores all instances of induced earthquakes in the Netherlands \citep{KNMI}. NAM has data on everything related to drilling and production. Namely, pore pressure measurements in the gas field can be found on NAM's website \citep{NAM}. Coordinates of the wells are available in the production plan 2003 \citep{GW03}. Finally, the monthly production values can be obtained from NAM representatives. 

\begin{figure}
    \centering
    \includegraphics[width=0.45\linewidth]{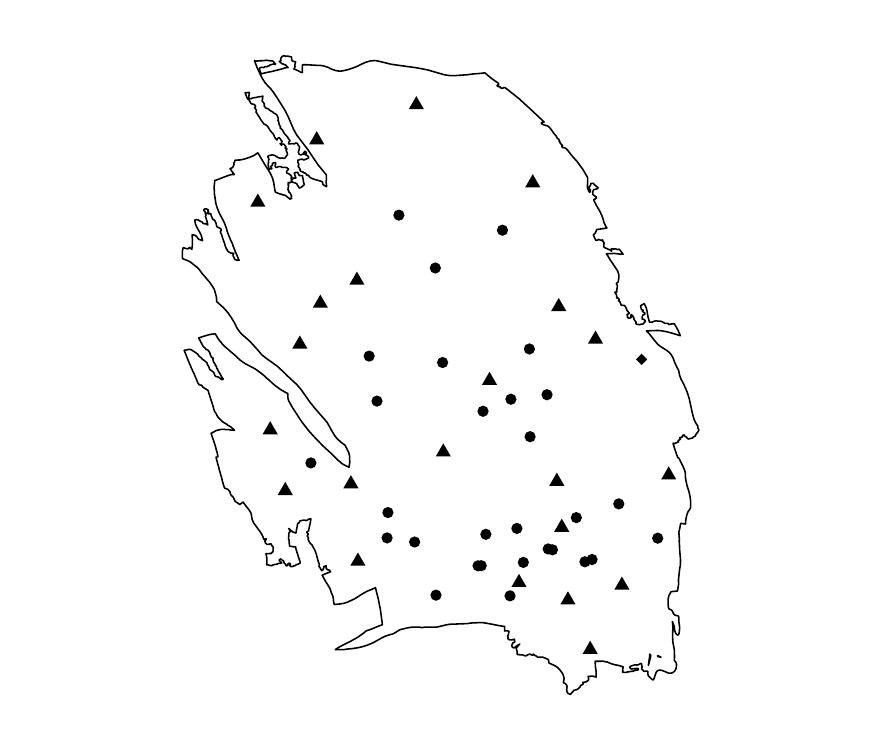}
    \includegraphics[width=0.45\linewidth]{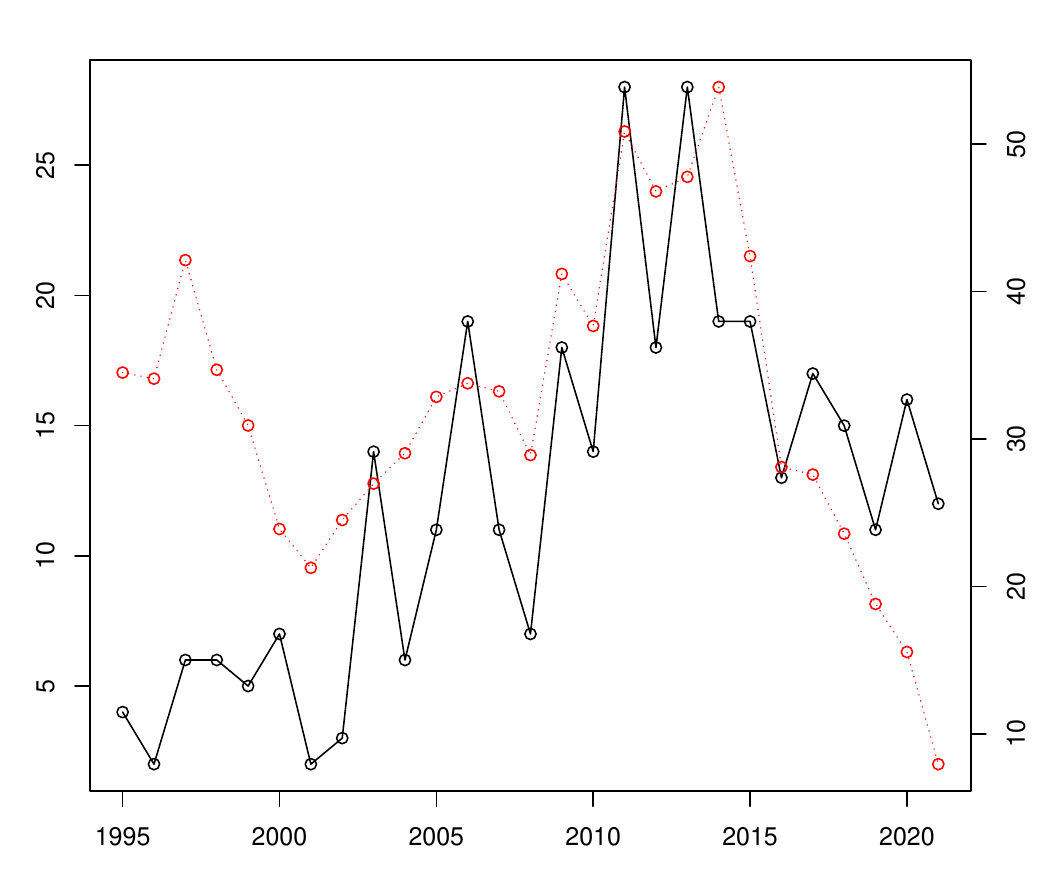}
    \caption{Left: Plot of the Groningen gas field with the drilled borehole: production (circles), observation (triangles) and injection (rhombus) locations. Right: Annual counts of the earthquakes and the production volumes in bcm.}
    \label{fig:fieldmap}
\end{figure}

Due to different data sources, there are inconsistencies in the coordinates data. TNO records the Groningen map using the UTM-31 zone, while KNMI and NAM record well and earthquake coordinates in the Amersfoort RD system. For consistency, we convert all locations' data into UTM-31 zones and restrict data points to locations within the gas field. The map of the borehole locations in the Groningen gas field can be seen in the left-hand-side of figure~\ref{fig:fieldmap}.

Data collection starting points are also different for each data source. Pressure observations are recorded haphazardly at different wells starting from April of 1960. Gas production figures go all the way up to 1956. The first instances of induced earthquakes were recorded in 1986. However, \citet{Dost12} report that due to the inaccuracy of the equipment used in the early days, induced earthquakes can be reliably recorded only starting from 1995. For the same reason, a magnitude threshold of $1.5$ or higher is also advised. Thus, we fixed January of 1995 as the starting point for our data and implemented a magnitude threshold $M\geq 1.5$.

Monthly gas extraction volumes are provided at each producing well in normal cubic meters (ncm). We rescale them into billion cubic meters (bcm) for convenience of calculations. The values are then smoothed throughout the spatial locations within the gas field, according to \citet{LieshBaki23}. The plot of annual production volumes and the earthquake counts is shown in the right-hand-side of figure~\ref{fig:fieldmap}.

Finally, NAM provides pore pressure estimates from the Groningen dynamic reservoir model. The model description and consecutive updates can also be found on NAM's web page. The pore pressure values are provided annually in Amersfoort RD coordinates within 500m grids. The coordinates are converted into the UTM-31 zone and filtered within the range of the gas field. In the next section, we describe the reservoir modelling process in more detail.

\section{Groningen reservoir model}\label{S:GFR}

During the drilling process, drilling and wireline logs are saved for further analysis by a team of geologists. These logs are then used to construct a map for a sequence of rock layers (rock strata) in the study area.   Once the lithostratigraphy is analysed and the field area is divided into sub-regions based on the core layers, reservoir properties must be assigned to predefined grid cells to construct the static model. This is done based on the geostatistical techniques. The wireline logs are interpolated for each given reservoir property (porosity, permeability, saturation, etc.) to derive the mean values for each grid crossed by the drilling trajectory. Vertical and horizontal trends are identified based on the resulting observations, and variograms are derived. The algorithm then creates values by honouring the log values for cells crossed by drill trajectories and simulating values for the rest of the grids based on the structure of the variograms \citep{NJG17_2}. The model can output properties such as clay percentage, porosity, permeability and water saturation. Once the static model is completed, the history-matching procedure for the dynamic model can be performed.  

The dynamic reservoir model uses four types of data to perform history matching. The static pressure (temperature) gradient (SP(T)G) data is routinely measured in the production and observation wells. It represents the increase in pressure per unit depth. Repeat formation tests (RFT) provide the measurement of pressure as a function of depth shortly after the drilling. There are only 41 RFTs available at Groningen \citep{NJG17}. Pulsed Neutron Log (PNL) data is another source for history matching in the field. It is a logging tool that provides information about the formation's porosity and fluid content. 217 pulsed neutron log surveys in 30 wells have been conducted in Groningen by 2016 \citep{NAMJun16}.  Finally, subsidence data are available from satellite and levelling surveys \citep{NJG17}.  It is used as a proxy for compaction,  which is harder to measure. Subsidence data proves to be especially useful in the north and north-western areas of the field where wells are scarce, and fewer pressure observations are available. 

The main focus of the dynamic model is the pore pressure since the pressure depletion is the root cause of induced seismicity in Groningen.  The reservoir pressure is modelled as a function of cumulative offtake and gas deviation factor:
\begin{align}\label{eq:CGR1}
    CGR_{cum}(s,t) = \frac{N_P(s,t)}{G_P(s,t)}
\end{align}
where $CGR_{cum}$ is the cumulative condensate gas  ratio (CGR), $N_P$ is the cumulative condensate production, and $G_P$ is the cumulative gas production. There is a well-established trend for the cumulative condensate gas  ratio as a function of pressure over gas expansion factor $P/z$:
\begin{align}
    CGR_{cum}(s,t) = c_1 \frac{P(s,t)}{z(P,T)} +c_2
\end{align}
where $c_1$ and $c_2$ are fitting constants to historical data, $P(s,t)$ is the field average pressure as a function of time and $z(P, T)$ is the gas expansion factor as a function of field average pressure and temperature, see \citet{NAMJun16}. However, the cumulative CGR cannot be used for forecasting; thus, the instantaneous CGR is used instead
\begin{align}
    CGR(s,t) = \frac{N_P(s,t) - N_P(s,t-\Delta t)}{G_P(s,t) - G_P(s,t-\Delta t)}.
\end{align}
for $\Delta t$ discrete time step.  Thus, the equation for pressure match becomes:
\begin{align}
    P(s,t) = \frac{z(P,T)}{c_1} \Big[ \frac{N_P(s,t) - N_P(s,t-\Delta t)}{G_P(s,t) - G_P(s,t-\Delta t)} - c_2 \Big]
\end{align}
The smoothness of the result obtained by the above formula depends on the time step length. The GFR uses annual time steps for yearly planning purposes.  

Given a pressure change $\Delta P$, the numerical calculation of subsidence above a compacting reservoir is given as:
\begin{align}
    u_z(x,y,0) = \frac{1-\nu}{\pi} \sum_{n=1}^N c_{mn}\Delta P_n \frac{L_{zn} l_{xn} l_{yn} l_{zn}}{[ (x-L_{xn})^2 + (y-L_{yn})^2 +L_{zn}^2]^{1.5}}.
\end{align}
for $u_z(x,y,0)$ subsidence (in m) above surface coordinates $x,y$. Here, $\nu$ is a Poisson ratio, $c_{mn}$ is the uniaxial compressibility of each grid block, $L_{xn}, L_{yn}, L_{zn}$ are $x,y,z$ directional distances of $l_{xn},l_{yn},l_{zn}$ dimensioned block $n$ from surface location $(x,y,0)$. See Figure 38 in \citet{NAMJun16} for a better understanding.   

The goal of the reservoir model is to match the pressure and subsidence model outputs to available data from the Groningen field and peripheral Land Assets in order to use the model for forward predictions. The workflow starts with a team of specialists determining the uncertain model parameters that are likely to impact the mismatch in data. The parameter ranges are based on realistic bounds and physical characteristics. Given the large number of parameters, the experimental designs might require a large number of simulations. Simulations are performed a user-specified number of times, and parameters are sampled from a uniform distribution within specified ranges that are set symmetrically around their base rate values.
Residual mean squared error (RMSE) is used to quantify the difference between model output and the observed data. The average RMSEs over the entire field and local regions are calculated for each data set. The models with the lowest RMSE values are selected among the resulting simulations. Local RMSE values are compared to model parameters to improve local matches and find any correlations. If correlations are found, then the parameter is manually set to its optimal value to obtain minimum local RMSE. For examples, see \citet{NJG17}. When the model results in local inconsistencies, the subsurface team reviews the set of parameters and makes necessary adjustments. See examples in \citet{NJG17}. The process is repeated until the match is consistent and minimal global and local RMSE is achieved.

Several Shell and NAM in-house software, including Petrel, MoReS, and Reduce++, were used in the process. The workflow provides annual estimates of pore pressure values in a user-specified grid size. NAM's first comprehensive reservoir model was built in 2012 to support the production plan. The history match is performed there based on SP(T)G, RFT and PNL data. It has been reviewed by an independent consultant, SGS Horizon and their suggestions were taken into account in later updates.  Several updates have been performed since then to ensure a continuous history match. A short summary of the main updates to the model over the years can be found in table~\ref{tab:upds} in the Appendix. The last results from the reservoir model were published in August of 2023. It takes into account the closing date of October 1st, 2023.

\section{Modelling and prediction of earthquake events}\label{S:model}
\subsection{Methodology}
As discussed previously, we assume the earthquakes $\Phi$ to constitute a Cox process in the spatio-temporal region $W_S\times W_T$ with intensity measure $\Lambda$:
\begin{align*}
    \Lambda(s, t_0 + k\Delta) = \frac{\gamma_0\exp\{ \theta_1 + \theta_2 V(s,t_0+k\Delta)\}}{\Gamma(s,t_0+k\Delta)}.
\end{align*}
Thus, we can think of the number of earthquakes $N(s,t)$ at cell $(s,t)$ to be Poisson distributed with mean $\Lambda(s,t)\Delta\Delta_S$ for cell area $\Delta_S$. Reparametrising $e^\beta = \alpha/\gamma_0$, we rewrite the intensity as
\begin{align}
   \Lambda(s,t_0+k\Delta) = \frac{\exp\{ \theta_1 + \theta_2 V(s,t_0+k\Delta)\}}{S_P(s,t_0+k\Delta)}
\end{align}
for
\begin{align}
    S_P(s,t_0+k\Delta) & = \exp\{ \alpha [P(s,t_0+k\Delta)- P(s,t_0))]\} \nonumber \\ &+ e^\beta\Delta \sum_{i=0}^{k-1} \exp\{ \alpha [P(s,t_0+k\Delta)- P(s, t_0 +i\Delta)\}].
\end{align}
The pressure $P(s,t)$ here is a sum of a function $m(s,t)$ that closely approximates the observed pressure values, such that $m(s,t) = \oE\{P(s,t)\}$ and a Gaussian random error $E(s,t)\sim N(0,\sigma^2)$. In our previous work \citep{BakiLies24}, we have fitted a polynomial function to the observed pressure measurements (SP(T)G data) to obtain the function $m(s,t)$. Here, we propose to use the outputted pressure estimates from NAM's reservoir model. 

The final update to the reservoir model \citep{NAMFin} provides annual estimates of pore pressure within 500m grid cells and RMSE values for each dataset used in the history matching procedure. According to NAM representative Jan van Elk: "the SP(T)G data is the most extensive pressure data set, measured in many cluster and observation wells across the field, and over the entire lifetime. It has the best coverage of reservoir pressure in both space and time. Thus, regarding how well the model can be expected to match pressure measurements in Groningen wells, the SP(T)G RMSE value is most appropriate." Thus, our random component will use the SP(T)G RMSE value of $2.3$ as a standard deviation.

Keeping to the same notation, we denote $\tilde{m}(s,t)$ to be estimates of pore pressure from the dynamic reservoir model, with tilde sign signifying a finer spatial grid of $500$m. According to the NAM representatives, an arithmetic mean is sufficient enough for downscaling to a larger grid size. Thus, to match our chosen gird size of $1$km, we define the new deterministic function
\begin{align*}
    m(s,t) = \frac{1}{4}\sum_{\tilde{s}} \tilde{m}(\tilde{s},t) \quad \mbox{such as} \quad d(\tilde{s},s)\leq 0.5
\end{align*}
for an $l_1$ distance function $d()$ and spatial locations $s$. The assumption is that the true pressure value at cell $(\tilde{s},t)$ is Gaussian distributed with mean $\tilde{m}(\tilde{s},t)$ and variance $2.3^2 = 5.29$. Thus it is straightforward that the variance of $E(s,t)$ within $1$km discretisation is $\sigma^2 = \frac{1}{4^2} (4\times RMSE) = 1.3225$ for an $m(s,t)$ function above. One might argue that less than four locations $\tilde{s}$ might fall within the downscaled grids $s\in W_S$, especially along the borders of the gas field. However, since the model intensity is already scaled by the grid area within the gas field, we choose to keep a constant variance for all cells.   
  
Following our previous work \citep{BakiLies24}, we use the first-moment measure of the Cox process to estimate the parameters. Then, run a Metropolis Adjusted Langevin Algorithms (MALA) for the posterior distribution of $E(s,t)|N(s,t)$ to monitor earthquake intensity in the next time interval. For details on methodology, see \citet{BakiLies24}. 

\subsection{Estimation and monitoring}
An estimating equation specified in \citet{BakiLies24}, with $1$km grids and annual temporal discretisation yields the following parameter estimates: $\hat{\theta}_1 = -5.47$, $\hat{\theta_2} = 13.32$, $\hat{\alpha} = 0.0129$ and $\hat{\beta} = -16.28$. Here $\theta_1$ is an intercept, and $\theta_2$ quantifies the effect of changing production volumes. Since $e^{\hat{\beta}} = 8.5\times 10^{-8}$, the model can be simplified to 
\begin{align*}
    \Lambda(s,t) = \exp\{\theta_1 + \theta_2 V(s,t) +\alpha [P(s,t_0) - P(s,t)] \}.
\end{align*}
Annual gas production figures were rescaled into billion cubic meters for convenience of calculations. Thus, assuming a constant pressure, an increase in production by a million cubic meters will increase the intensity rate by approximately $1.34\%$. Whereas a drop in pressure by one unit with constant production will increase the intensity by $1.29\%$. The production in the Groningen gas field stopped in October of 2023. Therefore, if we assume an extreme long-run scenario of $P(s, T)$ approaching zero at some time $T$, given that the maximum pressure recorded was $354$ barA, the earthquake intensity will approach approximately $0.39$, making the probability of one or more event occurring $\oP(n>0) = 0.3$. However, it is known that due to the fluid flow, the pressure in the reservoir will stabilise at some equilibrium value and, in certain areas, might increase, leading to smaller intensity estimates.        

As before \citep{BakiLies24}, we have used the Godambe information matrix to estimate $95\%$ confidence intervals for parameter estimates. They are $(-5.67, -5.27)$ for $\theta_1$, $(9.2, 17.44)$ for $\theta_2$ and $(0.0096, 0.0161)$ for $\alpha$. For comparison, a parameter fitting procedure with the polynomial function $m(s,t)$ fitted to the observed pressure values \citep{BakiLies24} provided estimates of $\hat{\theta}_1 = -5.33$, $\hat{\theta}_2 = 14.56$, $\hat{\alpha} = 0.01$ and $\hat{\beta} = -17.4$ with confidence intervals $\theta_1\in (-5.52, -5.14)$, $\theta_2\in(10.42,18.7)$ and $\alpha\in(0.007, 0.013)$. As one can observe, the estimates from both models fall within each other's confidence intervals, which confirms the estimates once more.   

\begin{figure}
    \centering
    \includegraphics[width = 0.45\linewidth]{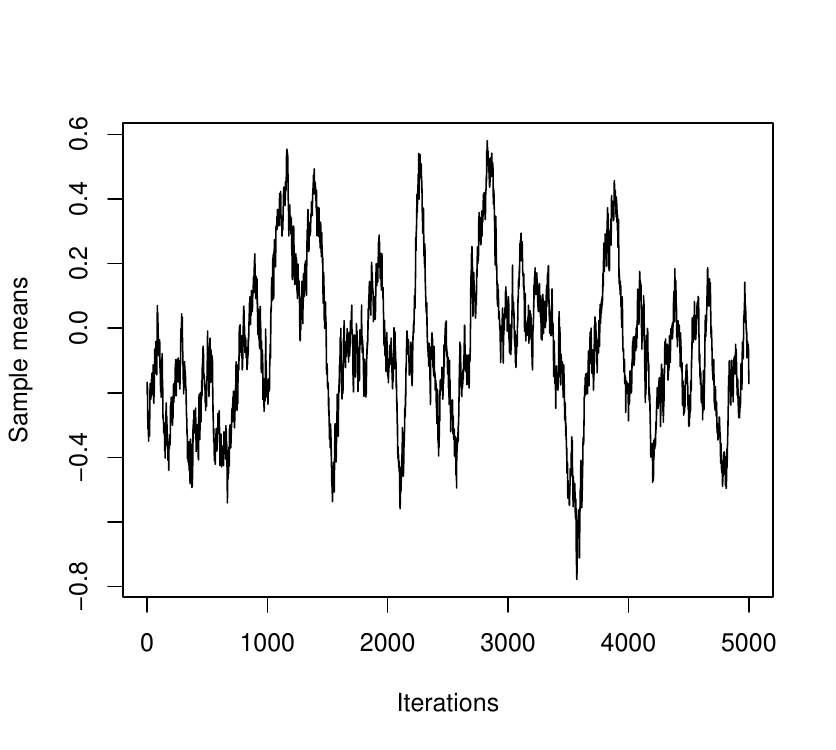}
    \includegraphics[width = 0.45\linewidth]{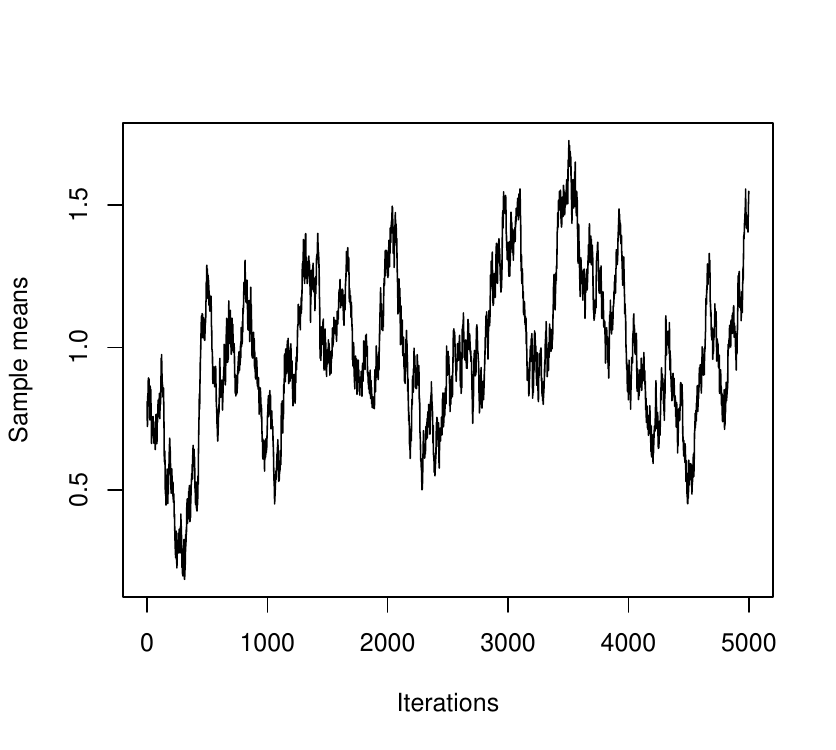}
    \caption{Trace plots of sample means $\sum_{t=1}^T E(s,t)/N_T$ at two randomly chosen locations $s$.}
    \label{fig:samp_means}
\end{figure}

To analyze the convergence and mixing of the MCMC algorithm, we use trace plots. Figure~\ref{fig:samp_means} displays the mean sampled values of our parameter interest $E(s,t)$ over the algorithm iterations. Since the algorithm performs sampling at each location individually, the plots also show trace plots for two randomly selected locations $s\in W_S$. 
We first used trace plots to identify the possible number of iterations for the burn-in period. Since the variance is quite small, one could assume that a smaller number of iterations are needed to achieve convergence. However, analysis of the trace plots shows that visible convergence and mixing occur only after 4000 iterations. Thus, we discard the first 5000 iterations as burn-in and perform a further 5000 iterations for sampling. As one can see from figure~\ref{fig:samp_means}, the parameter values fluctuate within a consistent range without any apparent trend, suggesting the chain has converged to the target posterior distribution. An absence of an apparent trend also indicates good mixing and that the chain has reached the stationary distribution.

\begin{figure}[ht]
    \centering
    \includegraphics[width = 0.45\linewidth]{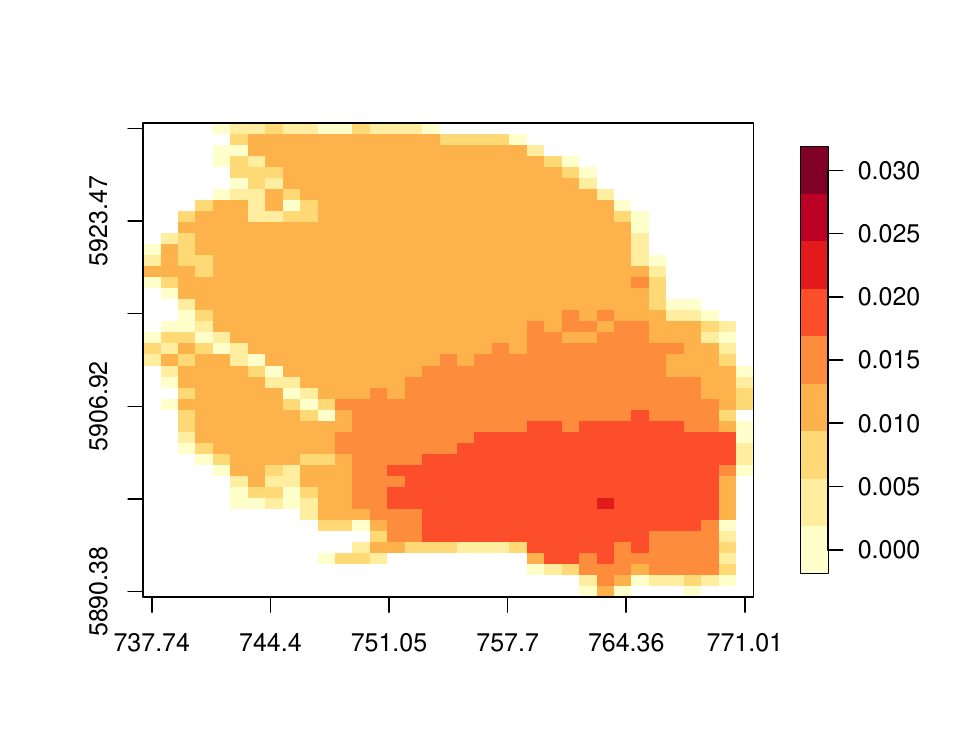}
    \includegraphics[width = 0.45\linewidth]{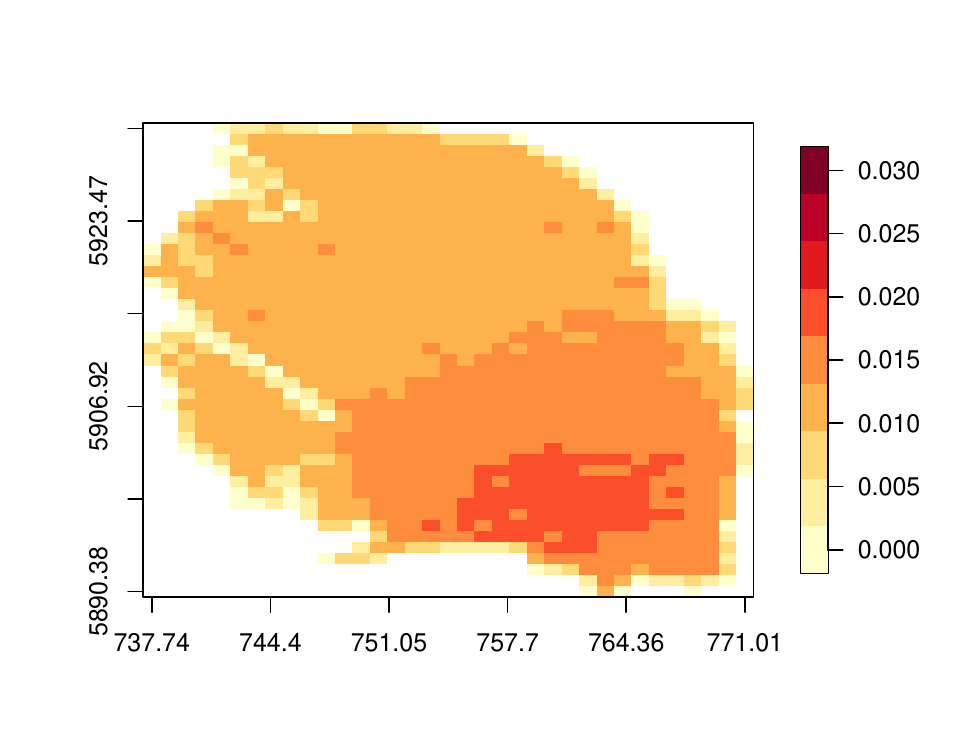}\\
    \includegraphics[width = 0.45\linewidth]{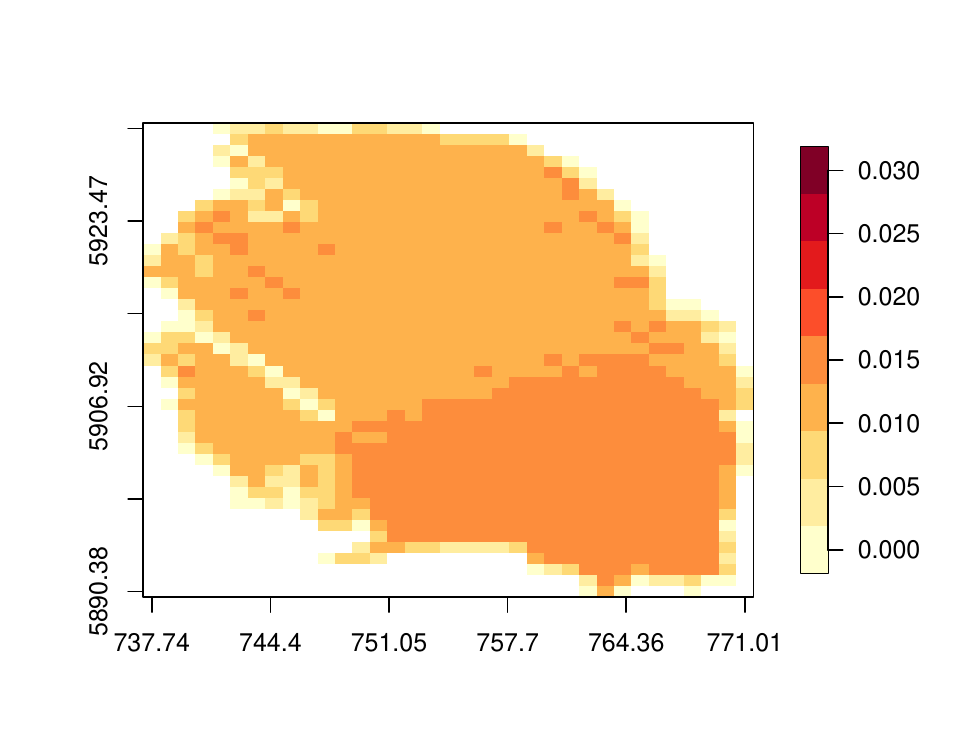}
    \includegraphics[width = 0.45\linewidth]{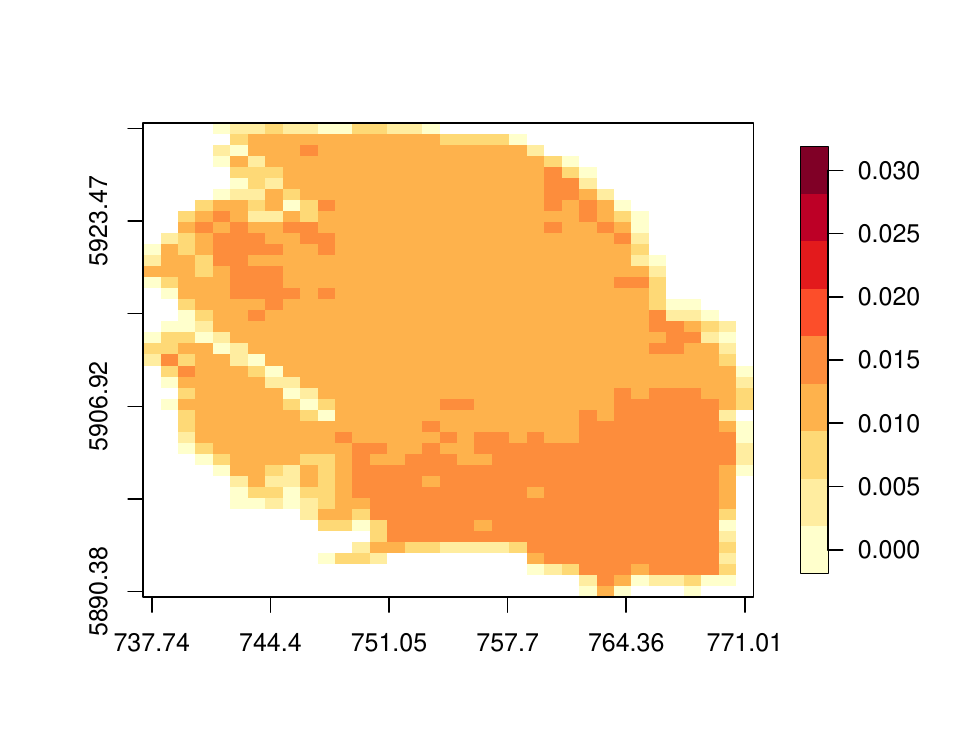}
    \caption{Predicted intensity plots over 1 km grid within the Groningen gas field. The graphs correspond to the years 2022 (top-left), 2023 (top-right), 2024 (bottom-left) and 2025 (bottom-right). XY coordinates represent UTM-31 coordinates.}
    \label{fig:pred_int}
\end{figure}
Finally, figure~\ref{fig:pred_int} shows predicted intensities for the model at hand between the years 2022 and 2025. Our analysis used data from 1995 to 2021; thus, 2022 is the first estimation period. Additionally, since the model depends on production volumes in a year preceding time $t$, the prediction is extended up to 2025 to show the effect of the absence of production in 2024. The figure shows mean intensity values drawn from 100 samples each year. As one can see, in 2022 and 2023, higher intensity values correspond to the south and southeast regions of the gas field, where the majority of the production wells are located. However, as we progress to the years 2024 and 2025, we see this apparent trend disappearing due to gas field closure because, according to the model, without production, the intensity of the earthquakes depends only on the pore pressure changes.

An increased number of production wells in the south of the Groningen gas field also results in lower pore pressure in the area. Thus, increasing the value of pressure drop $P(s,t_0) - P(s,t)$ and further contributing towards higher earthquake intensities. However, as geologists suggest, the pressure would move towards a new equilibrium state in the absence of production due to fluid redistribution. This would involve the migration of fluids from higher-pressure areas into the depleted reservoir areas, leading to an increase and/or decrease in pressure. The pressure recovery rate can vary significantly depending on the permeability and connectivity of the reservoir and surrounding formations. Thus, in the long run, the pore pressure in the reservoir and, consequently, the earthquake intensity will approach a new equilibrium state that is influenced by the surrounding geological features and the initial conditions prior to production. The final equilibrium pressure may not return to the original pre-production levels, especially if significant deformation or compaction has occurred. However, one would expect a decrease in intensity associated with the drop in pressure changes and an eventual stabilization at some small value. This would also entail a lesser spatial variability influenced only by the rock formation structure, as can be seen in the figure~\ref{fig:pred_int} map of 2025.

\subsection{Discussions}
We have constructed a Cox rate-state model using historical data to predict earthquake intensities in the Groningen gas field, where the intensity is a function of gas production and pore pressure drop. Two different approaches were employed to tackle the issue of sparse pore pressure observations. First,  a polynomial function is fitted to the observed data in \citet{BakiLies24}. This approach is relatively straightforward to implement and easily interpretable.
It is purely data-driven since the method uses observed data and can capture specific trends and anomalies in historical pressure measurements. Since polynomials can be adjusted to fit the data as closely as necessary by increasing and/or decreasing the order, it provides a high degree of flexibility.
However, depending on the order of the polynomial, it can overfit the historical data, capturing noise rather than the underlying trend and leading to poor generalization to future data points. This might result in unpredictable scenarios outside the range of observed data, leading to unreliable predictions for unobserved spatio-temporal points.

A second approach explored in this paper uses the output of the NAM's reservoir dynamic model. It is more theoretically grounded since the reservoir model uses geomechanical equations that incorporate the physical principles governing pressure changes in the reservoir. By incorporating field measurements and laboratory experiments, the reservoir model can potentially provide more accurate predictions, especially in unobserved regions, due to its basis in physical laws. Additionally, the model output typically has a smaller variance because it smooths out the noise present in historical data, focusing on the fundamental physical processes. However, the accuracy of the model heavily depends on the quality and precision of parameter estimates, and thus, inaccurate field measurements or laboratory data can lead to significant errors in predictions. As described in the history matching workflow \citep{NJG17}, among many simulated samples, a model with the lowest local and global RMSE value is chosen. Thus, since the goal of the workflow is to match the historical data best, the model has lower variance and might potentially miss some realistic variability.  

\begin{figure}[ht]
    \centering
    \includegraphics[width = 0.45\linewidth]{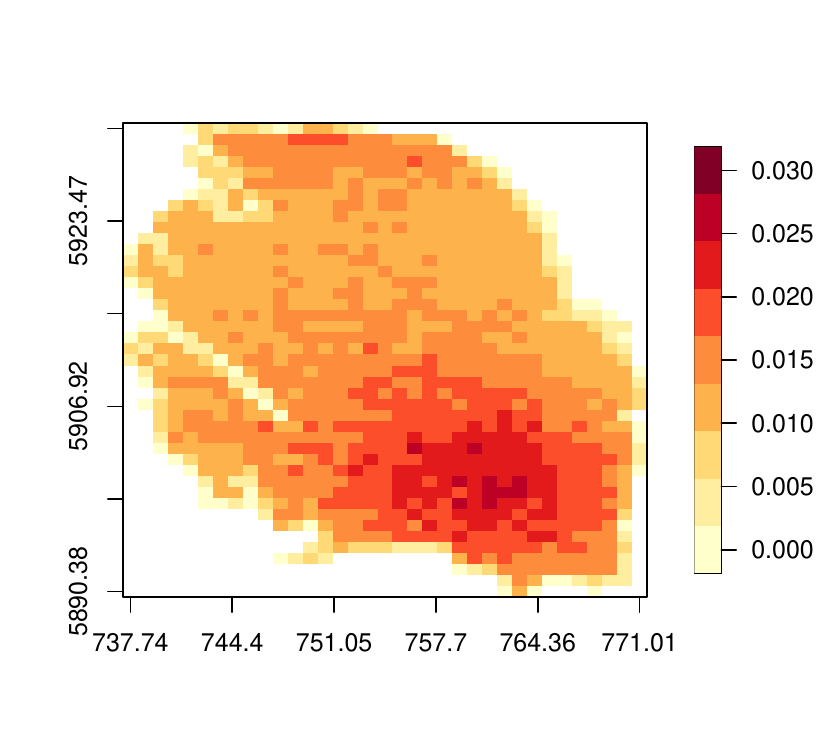}
    \includegraphics[width = 0.45\linewidth]{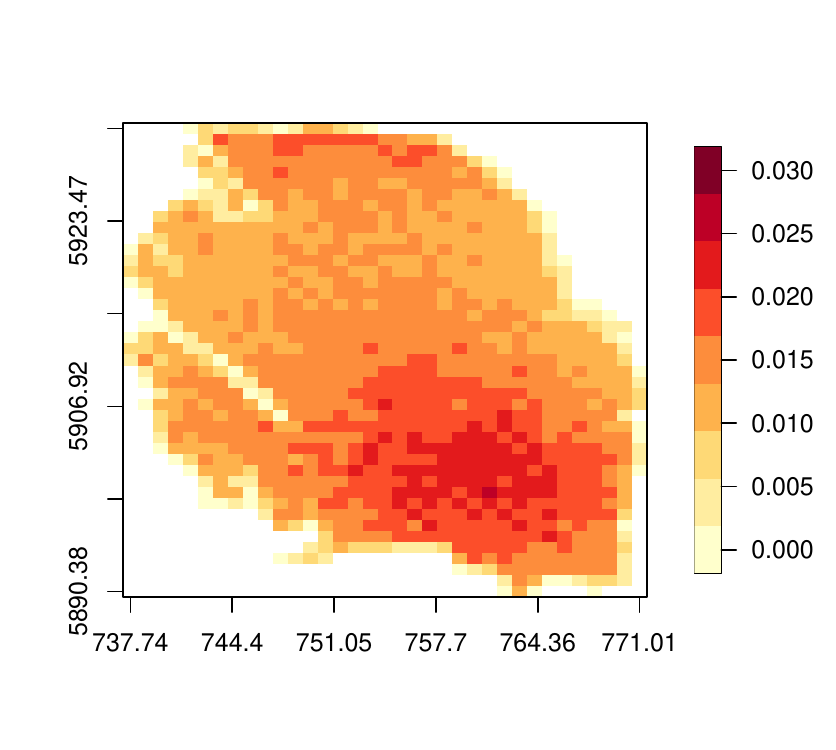}\\
    \includegraphics[width = 0.45\linewidth]{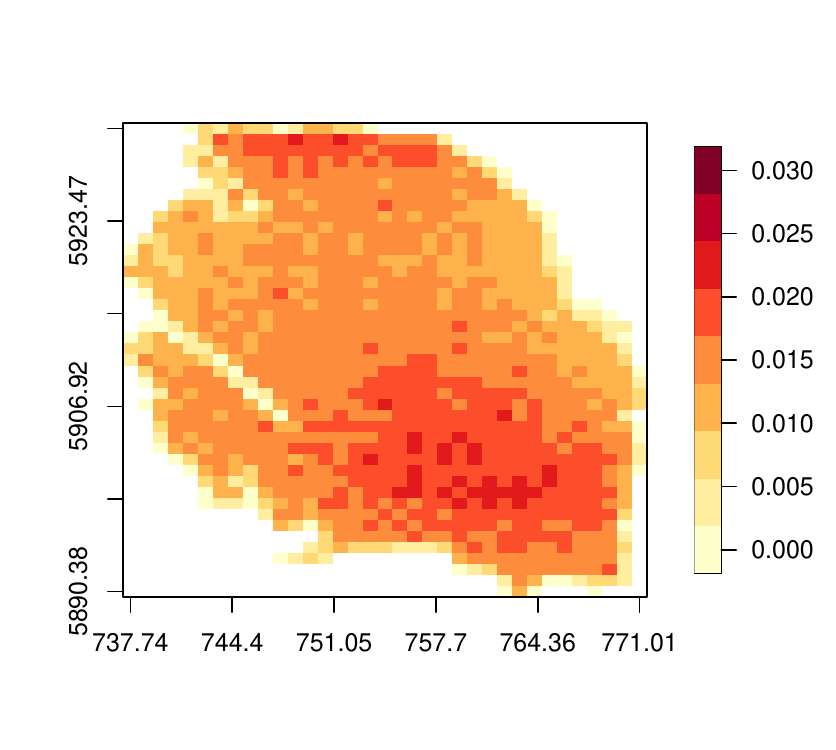}
    \includegraphics[width = 0.45\linewidth]{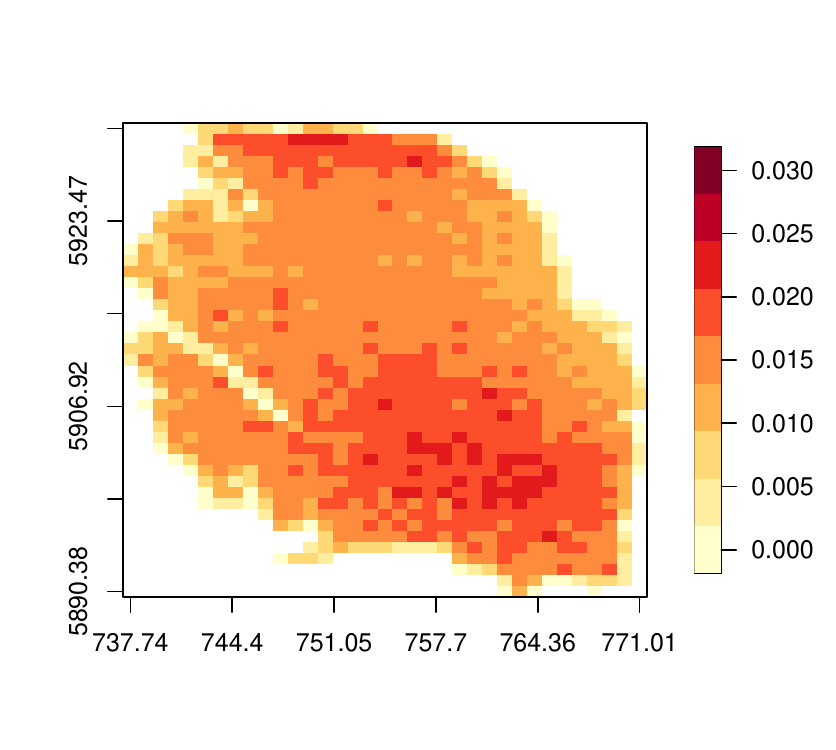}
    \caption{Predicted intensity plots from \citet{BakiLies24}, over 1 km grid within the Groningen gas field. The graphs correspond to the years 2022 (top-left), 2023 (top-right), 2024 (bottom-left) and 2025 (bottom-right). XY coordinates represent UTM-31 coordinates.}
    \label{fig:pred_int_old}
\end{figure}

For comparison, we have plotted the mean estimated intensity maps from the polynomial model in \citet{BakiLies24} in figure~\ref{fig:pred_int_old}. It also has higher intensity values in the south of the gas field. Additionally, since the polynomial $m(s,t)$ was fitted to the observed pressure data between 1995 and 2021, it continues in the trajectory of decreasing pore pressure due to gas extraction, having larger intensity estimates in the years 2024 and 2025. However, induced earthquakes recorded in recent years (2022-2023) have occurred more in the central and northwestern areas of the gas field. Thus, the model does well in capturing such a trend. For comparison, the NAM model shows a visible increase in intensity in the northern regions only by 2025, focusing more on the production areas and pressure behaviour. We believe this must be due to the narrow estimates of the NAM's model, which fails to capture some of the underlying realistic variability. In fact, some geology and seismology experts in the field have also voiced their concerns about the overconfidence of the NAM's reservoir model, which is missing a realistic scenario. 

To conclude, both models perform well in capturing the general spatio-temporal trends in induced earthquake intensity due to gas extraction and pore pressure drop. Polynomial $m(s,t)$ model from \citet{BakiLies24} better captures the spatial variability of earthquake intensity and observed earthquake locations in recent years. However, for better predictions, a recalibration of the polynomial function is needed to capture the pressure behaviour after 2021, especially once production has stopped. The NAM's reservoir model output analysed in this paper takes into account the entire history of the gas field up to the moment of closure. It greatly describes the pressure behaviour in the gas field, yet it fails to account for the spatial variability of observed events stemming from the unrealistically small variance of the proposed model output.        

\section{Conclusion}\label{S:conc}
This paper underscores the importance of accurately modelling pore pressure changes to understand and predict induced seismicity in the Groningen gas field. By integrating NAM's dynamic reservoir model data with a Cox process-based rate-state model, we have provided a comprehensive approach that combines geomechanical insights with statistical rigour. The model parameters estimated through MCMC algorithms indicate a robust relationship between pressure depletion and seismic intensity, confirming the critical role of pressure management in mitigating seismic hazards.

Our findings suggest that future seismic activity can be effectively monitored and predicted using our integrated modelling approach. The results emphasize the need for continuous updates and refinements of the reservoir model to ensure accuracy in reflecting real-time pressure changes. As the Groningen gas field transitions away from active production, our model offers a valuable tool for assessing long-term seismic risks, contributing to safer and more sustainable resource management practices.

\section*{Acknowledgements}

The Dutch Research Council NWO funded this research through their DeepNL programme (grant number DEEP.NL.2018.033). We are grateful to Mr Van Eijs and Ms Landman for providing us with data on gas production and useful insides on the Groningen reservoir model.

\newpage
\appendix
\section{Reservoir model updates}
\begin{table}[ht!]
\centering
\begin{tabular}{|c|p{0.8\textwidth}|}
\hline
\textbf{Version} & \textbf{Updates} \\ \hline
GFR 2015 & \begin{itemize}[leftmargin=*, noitemsep, topsep=0pt, parsep=0pt, partopsep=0.5pt]
    \item Additional medium of subsidence is added to the history matching; 
    \item Existing grid boundaries extended $8-10$km to the West and $5$km to the South \citep{NAMJun16}; 
    \end{itemize}\vspace{-\topsep} \\
    \hline
GFR 2017 & 
\begin{itemize}[leftmargin=*, noitemsep, topsep=0pt, parsep=0pt, partopsep=0.2pt]
     \item Closed-in tubing-head pressures converted to bottom-hole pressures (CITHP2BHP) data was added;
    \item Time-lapse gravity data was added for subsidence data;
    \item Interpolation of porosity between well locations was updated in the static model \citep{NAMSep17};
    \end{itemize}\vspace{-\topsep} \\
    \hline
GFR 2018 & 
\begin{itemize}[leftmargin=*, noitemsep, topsep=0pt, parsep=0pt, partopsep=0.2pt]
     \item Time period of the history matching data was stretched up to April of 2017;
     \item Updates from new iterations of seismic inversion and rock properties derived from Zeerijp-3A special core analysis were added to the static model \citep{NAMJun18};
    \end{itemize}\vspace{-\topsep} \\
    \hline
GFR 2019 & 
\begin{itemize}[leftmargin=*, noitemsep, topsep=0pt, parsep=0pt, partopsep=0.2pt]
    \item The history-matching timeframe up to December 2017;   
    \item Strain data from the Zeerijp-3A DSS cable was added;
    \item The effect of gas-in-aquifer on pressure drop, compaction and subsidence is added $\rightarrow$ Improved predictions of subsidence in the northwest of the field;
    \item Analysis of gas-bearing Carboniferous was added $\rightarrow$ Improved subsidence predictions in the south of the field; 
    \item The depletion in the Lauwerszee aquifer resulting from gas production was added \citep{NAMOct18};
    \end{itemize}\vspace{-\topsep} \\
    \hline
GFR 2023 & 
\begin{itemize}[leftmargin=*, noitemsep, topsep=0pt, parsep=0pt, partopsep=0.2pt]
    \item Corrections for Ten Boer and Ameland layer thicknesses in the static model was implemented $\rightarrow$ Improved local mismatches at wells BDM-5, DZL-1 and BRW  \citep{NAMFin};  
    \end{itemize}\vspace{-\topsep} \\
    \hline
\end{tabular}
\caption{Summary of updates to the Groningen reservoir model}
\label{tab:upds}
\end{table}


\newpage
\bibliographystyle{plainnat}  
\bibliography{bibliography}

\begin{thebibliography}{24}
\providecommand{\natexlab}[1]{#1}
\providecommand{\url}[1]{\texttt{#1}}
\expandafter\ifx\csname urlstyle\endcsname\relax
  \providecommand{\doi}[1]{doi: #1}\else
  \providecommand{\doi}{doi: \begingroup \urlstyle{rm}\Url}\fi

\bibitem[Baki and van Lieshout(2024)]{BakiLies24}
Z.~Baki and M.N.M. van Lieshout.
\newblock A cox rate-and-state model for monitoring seismic hazard in the groningen gas field.
\newblock \emph{arXiv:2403.13413v2}, 2024.

\bibitem[Broccardo et~al.(2017)Broccardo, Mignan, Wiemer, Stojadinovic, and Giardini]{Broc17}
M.~Broccardo, A.~Mignan, S.~Wiemer, B.~Stojadinovic, and D.~Giardini.
\newblock Hierarchical bayesian modelling of fluid-induced seismicity.
\newblock \emph{Geophysical Research Letters}, 44:\penalty0 11357--11367, 2017.

\bibitem[Burkitov et~al.(2016)Burkitov, van Oeveren, and Valvante]{NAMJun16}
U.~Burkitov, H.~van Oeveren, and P.~Valvante.
\newblock Groningen field review 2015 subsurface dynamic modelling report.
\newblock Technical report, NAM Report No. EP201603238100, 2016.

\bibitem[Candela et~al.(2019)Candela, Osinga, Ampuero, Wassing, Pluymaekers, Fokker, van Wees, de~Waal, and Muntendam-Bos]{Cand19}
T.~Candela, S.~Osinga, J.~Ampuero, B.~Wassing, M.~Pluymaekers, P>A. Fokker, J.~van Wees, H.A. de~Waal, and A.G. Muntendam-Bos.
\newblock Depletion-induced seismicity at the groningen gas field: Coulomb rate-and-state models including differential compaction effect.
\newblock \emph{Journal of Geophysical Research: Solid Earth}, 124:\penalty0 7081--7104, 2019.

\bibitem[Cox(1955)]{Cox55}
D.R. Cox.
\newblock Some statistical methods connected with series of events.
\newblock \emph{Journal of the Royal Statistical Society.}, 17(2):\penalty0 129--164, 1955.

\bibitem[de~Zeeuw and Geurtsen(2018{\natexlab{a}})]{NAMJun18}
Q.~de~Zeeuw and L.~Geurtsen.
\newblock Groningen dynamic model update 2018 - v5.
\newblock Technical report, NAM Report No. EP2018, 2018{\natexlab{a}}.

\bibitem[de~Zeeuw and Geurtsen(2018{\natexlab{b}})]{NAMOct18}
Q.~de~Zeeuw and L.~Geurtsen.
\newblock Groningen dynamic model update 2018 - v6.
\newblock Technical report, NAM Report No. EP201809202872, 2018{\natexlab{b}}.

\bibitem[Dempsey and Suckale(2017)]{DempSuck17}
D.~Dempsey and J.~Suckale.
\newblock Physics-based forecasting of induced seismicity at groningen gas field, the netherlands.
\newblock \emph{Geophysical Research Letters}, 22:\penalty0 7773--7782, 2017.

\bibitem[Dost et~al.(2012)Dost, Goutbeek, van Eck, and Kraaijpoel]{Dost12}
B.~Dost, F.~Goutbeek, T.~van Eck, and D.~Kraaijpoel.
\newblock Monitoring induced seismicity in the north of the netherlands: Status report 2010.
\newblock Technical report, KNMI, 2012.

\bibitem[Groningen Winningsplan()]{GW03}
Groningen Winningsplan, 2003.
\newblock Groningen Winningsplan 2003. Available at \url{https://www.nam.nl/oil-and-gas-production/groningen-gas-field/significance-of-the-production-plan/_jcr_content/root/main/section_998695721/promo_1349258685_cop/links/item0.stream/1672060188250/b00b60c900e4c00444a9c88a0cc4fa492018e4c8/groningen-winningsplan-2003-scanned1.pdf}.

\bibitem[Hajati et~al.(2015)Hajati, Langenbruch, and Shapiro]{Haja15}
T.~Hajati, C.~Langenbruch, and S.A. Shapiro.
\newblock A statistical model for seismic hazard assessment of hydraulic-fracturing-induced seismicity.
\newblock \emph{Geophysical Research Letters}, 42:\penalty0 601--606, 2015.

\bibitem[KNMI()]{KNMI}
KNMI, 2023.
\newblock Catalogue of induced earthquake data in the Netherlands from KNMI. Retrieved in April 2023. Available at \url{www.knmi.nl/kennis-en-datacentrum/datatset/aardbevingscatalogus}.

\bibitem[Landman and Visse(2023)]{NAMFin}
A.~Landman and C.~Visse.
\newblock Groningen dynamic model update 2023.
\newblock Technical report, NAM Report No. EP202306200914, 2023.

\bibitem[NAM()]{NAM}
NAM, 2023.
\newblock Pore pressure measurements at wells collected by NAM. Retrieved in April 2023. Available at \url{www.nam.nl/feiten-en cijfers/onderzoeksrapporten.html}.

\bibitem[Post et~al.(2021)Post, Michels, and Ampuero]{Post21}
R.A.J. Post, M.A.J. Michels, and J.P. Ampuero.
\newblock Interevent-time distribution and aftershock frequency in non-stationary induced seismicity.
\newblock \emph{Scientific Reports}, 11, 2021.

\bibitem[Richter et~al.(2020)Richter, Hainzl, Dahm, and Z\H{o}ller]{Rich20}
G.~Richter, S.~Hainzl, T.~Dahm, and G.~Z\H{o}ller.
\newblock Stress-based statistical modelling of the induced seismicity at the groningen gas field, the netherlands.
\newblock \emph{Environmental Earth Sciences}, 79, 2020.

\bibitem[Sijacic et~al.(2017)Sijacic, Pijpers, Nepveu, and van Thienen-Visser]{Sija17}
D.~Sijacic, F.~Pijpers, M.~Nepveu, and K.~van Thienen-Visser.
\newblock Statistical evidence on the effect of production changes on induced seismicity.
\newblock \emph{Netherlands Journal of Geosciences}, 96:\penalty0 27--38, 2017.

\bibitem[TNO()]{TNO}
TNO, 2023.
\newblock Geological coordinates of the Groningen gas field from TNO. Retrieved in April 2023. Available at \url{www.nlog.nl}.

\bibitem[Trampert et~al.(2022)Trampert, Benzi, and Toschi]{Tram22}
J.~Trampert, R.~Benzi, and F.~Toschi.
\newblock Implications of the statistics of seismicity recorded within the groningen gas field.
\newblock \emph{Netherlands Journal of Geosciences}, 101, 2022.

\bibitem[van Jaarsveld(2012)]{NAMJan12}
J.~van Jaarsveld.
\newblock Groningen field review 2012.
\newblock Technical report, NAM Report No. EP201202215894, 2012.

\bibitem[van Lieshout and Baki(2023)]{LieshBaki23}
M.N.M. van Lieshout and Z.~Baki.
\newblock Exploring seismic hazard in the groningen gas field using adaptive kernel smoothing.
\newblock \emph{Mathematical Geosciences: Special Issue}, 2023.

\bibitem[van Oeveren et~al.(2017{\natexlab{a}})van Oeveren, P., Geurtsen, and van Elk]{NJG17}
H.~van Oeveren, Valvatne P., L.~Geurtsen, and J.~van Elk.
\newblock History match of the groningen field dynamic reservoir model to subsidence data and conventional subsurface data.
\newblock \emph{Netherlands Journal of Geosciences}, 96(5):\penalty0 47--54, 2017{\natexlab{a}}.

\bibitem[van Oeveren et~al.(2017{\natexlab{b}})van Oeveren, Valvatne, and Geurtsen]{NAMSep17}
H.~van Oeveren, P.~Valvatne, and L.~Geurtsen.
\newblock Groningen dynamic model update 2017.
\newblock Technical report, NAM Report No. EP201708205454, 2017{\natexlab{b}}.

\bibitem[Visser and Solano~Viota(2017)]{NJG17_2}
C.~Visser and J.~Solano~Viota.
\newblock Introduction to the groningen static reservoir model.
\newblock \emph{Netherlands Journal of Geosciences — Geologie en Mijnbouw}, 96(5):\penalty0 39--46, 2017.

\end{thebibliography}
\end{document}